\documentclass[letterpaper, 10 pt, conference]{ieeeconf} 

\IEEEoverridecommandlockouts  

\usepackage{graphicx}

\title{\LARGE \bf
Influence of perspective taking through robotic virtual agents on prosocial behavior
}

\author{Chenlin Hang$^{1,3}$and Tetsuo Ono$^{2}$and Seiji Yamada$^{3,1}$
\thanks{This work was partially supported by JST, CREST (JPMJCR21D4),
Japan.}
\thanks{$^{1}$Department of Informatics, The Graduate University for Advanced Studies, SOKENDAI
        {\tt\small }}%
\thanks{$^{2}$Faculty of Information Science and Technology, Division of Computer Science and Information Technology, Hokkaido University
        {\tt\small }}%
\thanks{$^{3}$National Institute of Informatics
        {\tt\small }}%
}

\begin{document}

\maketitle
\thispagestyle{empty}
\pagestyle{empty}

\begin{abstract}
Perspective taking, which allows people to imagine another's thinking and goals, is known to be an effective method for promoting prosocial behaviors in human-computer interactions. However, most of the previous studies have focused on simulating human-human interactions in the real world by offering participants experiences related to various moral tasks through the use of human-like virtual agents. In this study, we investigated whether taking the perspective of a different robot in a robot-altruistic task would influence the social behaviors of participants in a dictator game. Our findings showed that participants who watched the help-receiver view exhibited more altruistic behaviors toward a robot than those who watched the help-provider view. We also found that, after watching robots from two different viewpoints in the task, participants did not change their behavior toward another participant.
\end{abstract}

\section{INTRODUCTION}
Technological artifacts such as autonomous cars, vacuum cleaners, smartphones, virtual agents, and robots that act autonomously in social environments are fast becoming a reality and are expected to increasingly interact with humans in a social way \cite{r27}. Many researchers have been fascinated with how these artifacts can persuade people to engage in prosocial behavior \cite{r17}\cite{r13}\cite{r23}\cite{r20}\cite{r22}\cite{r9}. Empathy, which plays a central role in human social relationships and is considered a major element in human social interaction, has been shown to increase understanding and motivate prosocial behaviors \cite{r3} \cite{r7} \cite{r6} \cite{r1}. Paiva et al. mentioned two possible systems for empathy: (1) a basic emotional and unconscious one and (2) a more advanced cognitive perspective-taking one \cite{r18}. As the basic system includes unconscious elements, which may lead to ethical problems, we focus on the perspective-taking system in this work. Extensive research has shown that taking the perspective of others (i.e., imagining what it would feel like to actually be the other person) can be a valid way of promoting prosocial behaviors \cite{r1}. Imagine-self perspective-taking tasks have typically resulted in better performances than the imagine-other approach for maximizing the motivation to help someone \cite{r8}. The image-self task, in which participants are instructed to imagine how they would feel if they were in someone else's situation, is somewhat different from the image-other task, in which participants imagine how someone else feels about a specific situation. With the development of technology, mediated perspective-taking tasks (e.g., online role playing games, videos, immersive virtual realities) that provide additional information to participants or users instead of relying only on the user's imagination have also shown good potential for promoting prosocial behaviors \cite{r12}\cite{r19}\cite{r2}. 

In the context of mediated perspective-taking tasks for promoting prosocial behaviors, most researchers have focused on simulating a specific situation based on a human-human interaction, and the virtual agents they use are almost always human-like in appearance \cite{r19}\cite{r2}\cite{r25}\cite{r15}. Considering the harmonious relationship between humans and robots, we need to examine whether taking the perspective of robots in a robot task can influence people's prosocial behaviors toward a human or a robot. Among the variety of elements that make up the prosocial task, in this work, we focus on altruism, which forms a major part of prosocial behavior and also plays a central role in our evolutionary origins, social relations, and societal organization \cite{r5}. On the basis of the definition of altruism, i.e., the behavior of a person who helps others at his or her own expense \cite{r4}, the altruistic task should contain at least two objects: the help provider and the help receiver. Therefore, we conducted an experiment to determine whether participants who took the perspective of different robots in a robot-altruistic task would be influenced in their social behaviors.

The between-participants experiment was conducted by showing different perspective views of robots engaged in a robot-altruistic task. One of the groups watched a help-provider-view video, and the other watched a help-receiver-view video. The task settings were based on a work by Hang et al. \cite{r28}, who applied nudge mechanisms in a video stimulus using robot-like virtual agents to promote human altruistic behaviors. They conducted a task in which two robots each had to arrange a meeting room, where one robot was lower-charged (treated as the receiver in our study) and the other was fully charged (treated as the provider in our study). The idea was that the provider would share its battery with the receiver when the receiver ran out of power. We measured behavior changes through a dictator game, which is a simple economic game typically used to measure altruistic attitudes. Our motivation was to determine not only the behavior changes toward a human but also toward a robot. The results showed that there was no difference in the behavior changes of participants toward a human. However, participants had significant differences in behavior change toward a robot after watching the different robot's perspective.

\begin{figure}[tb]
\centering
    \includegraphics[width=0.95\linewidth]{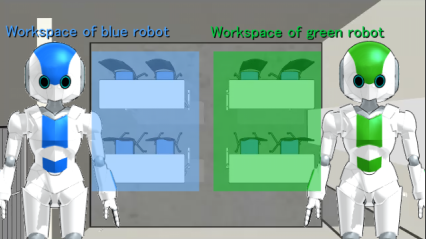}
\caption{Workspace of each robot}
\label{fig:verticalcell}
\end{figure}

\begin{figure}[tb]
\centering
    \includegraphics[width=\linewidth]{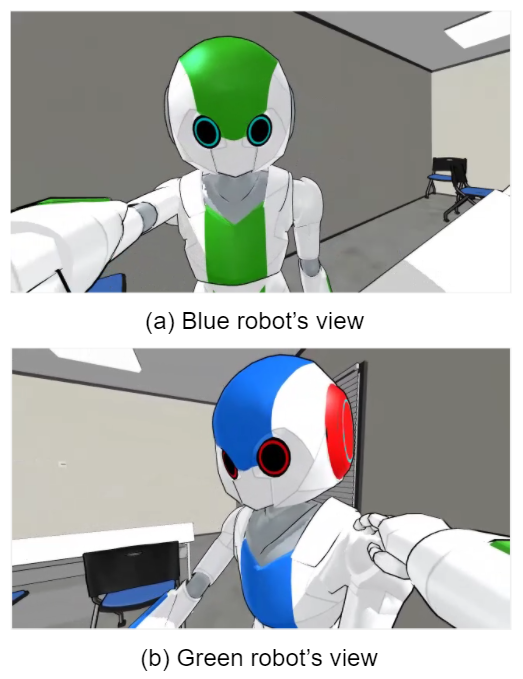}
\caption{The view of different robot in the same frame.}
\label{fig:verticalcell}
\end{figure}

\section{Related work}
\subsection{Prosocial behavior and altruism}
Prosocial behavior is a multidimensional concept that is broadly defined as voluntary behavior to benefit the other \cite{r29}\cite{r30}, such as altruism, solidarity, sharing, care-giving, and comforting \cite{r31}. There are also distinct types of prosocial behavior ranging from high-cost (e.g., extensive volunteering, helping in dangerous or emergency situations) to low-cost behaviors (e.g., helping to pick up a dropped item, sending an uplifting text message \cite{r30}). Based on the motivation behind prosocial behavior, including altruistic, cooperative, and individualistic behavior \cite{r32}\cite{r33}, altruism based on inter-reciprocity is a strong motivational factor in prosocial behavior \cite{r34}. 

Moreover, altruism is a central issue in our evolutionary origins, social relations, and societal organization \cite{r5}. There are three main categories of altruism: people who help others at their own expense \cite{r35}\cite{r4}, altruistic punishment \cite{r5}, and moral values \cite{r36}\cite{r37}. Among these three, in our current study, we focus on the definition of the unconditional prosocial tendency for an agent to benefit others without considering his/her own self-gain \cite{r35}.

\subsection{Altruism in human-robot interaction (HRI) and human-agent interaction (HAI)}
In the field of HRI, several robots have been developed to promote donation \cite{r38}, which has been treated as a representation of altruistic behavior \cite{r39}. In addition, in the fields of HRI and HAI, studies have been done on the factors of robots and agents that promote donation, such as the social background of users \cite{r40}, where a robot is installed \cite{r41}, the gender of a robot \cite{r42}, the appearance of a robot \cite{r43}, the head behavior of humanoid robots \cite{r44}, and the facial expressions of virtual agents \cite{r45}. 

Previous work has paid more attention to the characteristics of robots or agents and to the influence on the donation behavior of humans. Considering a possible hybrid society of humans and machines \cite{r46} in the near future, robots or agents may be members of human society. Nagataki et al. \cite{r47} asked participants to do a series of moral tasks after performing a bodily coordinated motion task with either another participant or a robot, and they observed that participants also made fair proposals to their robot partner. In our study, we would like to figure out whether experiencing a different view from the point of view of a robot in a robotic-altruistic task can influence the altruistic behavior of participants.

\subsection{Mediated perspective-taking}
Traditional perspective-taking tasks ask the participants imagine what it would be like to be somebody else under the specific circumstances. Different from that way, mediated perspective-taking tasks enable participants to get more information about the such circumstances through visual, auditory, and even haptic. Different forms of media (e.g., text, video, virtual reality, acted demo) have been used to convey interaction information vary in their level of immersion and interactivity. Although virtual reality performs best at immersion and interactivity, most psychological studies using VR systems based on laboratory settings where sample sizes are small and mostly composed of college students with little demographic variance\cite{r50}. The video stimuli have been shown to better perform in terms of social aspects according to the Almere model \cite{r49}. We set our experiment through the crowd sourcing service with video stimulus performed by robot-appearance virtual agents.

\section{Experiment}
\subsection{Video stimulus}
To imitate a real-life situation, Hang et al. \cite{r28} used virtual agents with a robot-like appearance in a video stimulus. On the basis of the definition of altruism (i.e., the behavior of a person who helps others at their own expense \cite{r4}), with the expense of the robots symbolized by their batteries, they investigated a scenario involving two robots doing the same task (organizing tables and chairs in a meeting room)(see Fig. 1.), where one of the robots suddenly stops working because its battery runs out. The battery of one robot was near 3\% while the other was fully charged, so the intent was to have the lower-charged robot stop working soon, and this would lead to altruistic behavior from the other robot. A beep sounded when the lower-charged robot's battery died, and its eyes and ears flashed a red light. After hearing the alarm and seeing the flashing lights, the fully charged robot moved towards the lower-charged robot and shared its battery power by placing a hand on its shoulder. 

In our study, we set the view of the lower-charged robot (blue robot) as the help-receiver's view and that of the fully charged robot (green robot) as the provider's view. We used MikuMikuDance, a content editing tool for 3DCG videos. To imitate the first-person perspective, we set the view of the video from the position of each robot's head. As there are two robots in the setting, we used one video for each robot. Fig. 2. shows the view from the two robots in the same frame, with that of the blue robot (lower-charged) on the left and that of the green robot (fully charged) on the right. 

\subsection{Participants}
Before collecting data for the experiment, we conducted a power analysis to determine the best sample size. A $G^\ast Power 3.1.9.7$ analysis \cite{r11} (effect size $f = 0.25$, $\alpha$ = 0.05, and 1 - $\beta$ = 0.80) suggested an initial target sample size of $N = 128$. A total of 176 participants (116 male, 60 female) ranging in age from 16 to 87 years ($M = 45.06$, $SD = 12.91$) took part in the experiment online. The participants were recruited through a crowd sourcing service provided by Yahoo! Japan. Regarding online experiments in general, Crump et al. \cite{r10} have suggested that data collected online using a web browser seem mostly in line with laboratory results, so long as the experiment methods are solid. 

Thirty-three participants were excluded due to a failure to answer comprehension questions on a video stimulus. Then, using the calculation of the $G^\ast Power 3.1.9.7$ analysis \cite{r11} as a basis, 64 participants were analyzed under each of the two conditions in our experiment in chronological order. The final sample of participants was composed of 128 (male = 84, female = 44, $M = 45.70$, $SD = 12.59$).

\subsection{Procedure}
We first asked the participants to read an introduction to the experiment and then watch a video stimulus. Next, to ensure they had watched the video completely, two comprehension questions were asked. One question was, ``What did the robots do in the video?", and the other was, ``What happened to one of the robots?" For each question, we provided four selections and asked participants to pick the correct answer. The selections to the first question ``What did the robots do in the video?" are "A.Cleaning, B.Walking, C.Preparing for the meeting, D.Guarding". The selections to the second question ``What happened to one of the robots?" are "A.The robot's leg was broken, B.The robot ran out of the battery, C.The smoke came out from the robot, D.The robot fell down". Participants were then asked to play the dictator game with two kinds of players (another participant and a robot) and state how much money they would give their fellow players if they had 1,000 yen. These two situations were shown in a random order to avoid the effect of order bias (i.e., practice effect \cite{r16}). Finally, a free-description question was asked to obtain comments from the participants at the end of the experiments. 

\subsection{Measurement}
We used the dictator game to measure the altruistic attitudes of participants as the dependent variable. In the basic structure of this game, S (the dictator) freely decides how much $x$ of an endowment to give to O (the recipient). O has no veto power; that is, O cannot react to S's decision\cite{r48}. On the basis of this structure, we set three different recipients (another participant, a robot). The reason we only chose the lower-charged robot was to forbid reciprocal thinking on the part of the green robot.

A real-life example of the dictator game would be one where players decide on making donations to a public charity, but considering that robots do not have any public charities, we set the questions on the basis of the basic structure. The questions were asked as follows. ``Now you are given 1,000 yen (about US\$8.70). The money needs to be shared with the other participant/robot. How much money would you give to the other participant/robot?" The answer was given using free input ranging from 0 to 1,000 yen.

\begin{figure}[tb]
\centering\includegraphics[width=0.5\textwidth]{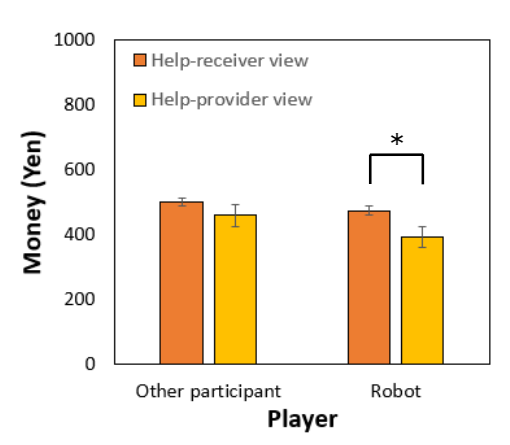}
\caption{The table of results } \label{fig4}
\end{figure}

\subsection{Results}
We analyzed the results of the dictator game for each player (the other participant, robot) separately with an independent sample $t$-test (see Fig. 3.).

The results for the other participant showed that there was no significant difference ($t(126)=1.83$, $p=0.07$) between the group that watched the help-provider's view and the group that watched the help-receiver's view. In contrast, the results for the robot showed that there was significant difference ($t(126)=2.20$, $p=0.029$) between the two groups. Moreover, participants who watched the help-receiver's view tended to give more money to the robot. 

In addition, we listed the typical comments among the free description part. There are totally 7 participants mention the empathy emotion after they finished video. The participants who watching help-receiver-view video mentioned more about the keywords, cooperation and altruism, than those who watching help-provider-view video. In contrast, the participants who watching help-provider-view video mentioned more about the crisis of the job replacement by robots(see Fig. 4.).

\begin{figure}[tb]
\centering\includegraphics[width=0.5\textwidth]{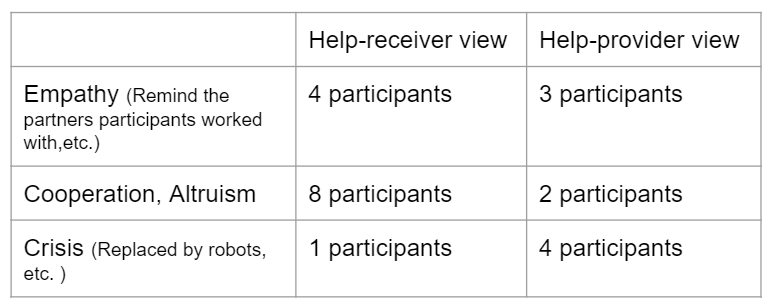}
\caption{The table of free descriptions } \label{fig7}
\end{figure}

\section{Discussion}
\subsection{General discussion}
We conducted this experiment to investigate whether experiencing two different points of view from virtual robots in the same altruistic task could promote human altruistic behavior. Our analysis of the data obtained from the experiment showed that, for the other participant, the participants did not exhibit much difference in their altruistic behavior after watching the videos of the help-receiver's view and the help-provider's view. 

Regarding the results for the other participant, although there was no significant difference in altruistic behavior after watching the help-receiver or help-provider’s video, the comments of the participants revealed that many of them felt a human-like quality in the robots when they performed altruistic behaviors (sharing battery power with the lower-charged robot), and this prompted them to recall their own coworkers or even to reflect on their daily behavior. Many previous works have shown that interacting with human-like agents actually enhances the prosocial behavior of participants \cite{r19}\cite{r2}\cite{r25}\cite{r15}. 

The significant difference between our work and previous research is not only the appearance of the virtual agent but also the task that is shown in the video stimulus. Therefore, it would be interesting to investigate the relationship between tasks and appearances that can not only promote human prosocial behavior but also deepen the relationship between human-robot and human-agent interactions. In addition, we plan to conduct additional experiments to determine whether there would be more of an influence when we use various types of equipment that enables participants to receive physical feedback. For example, by extending our video-based experiment to immersive virtual reality, participants could be made to really feel what it is like when the battery runs out, such as by making their body feel stuck.

The results for the robot showed that participants who watched the help-receiver-view video gave more money to the robot than those who watched the help-provider-view video. This demonstrates that taking the perspective of the help-receiver robot will cause people to give more money to a robot, which is similar to the experiment done by Baston et al. \cite{r7}, in which participants who took the perspective of a member of a stigmatized group reported more positive attitudes even toward the stigmatized group as a whole. Furthermore, comparing the free description after watching help-receiver-view video with help-provider-view video, the participants who watching help-receiver-view mentioned more about cooperation and altruistic.

Finally, the biggest difference between these two video stimuli was the period where the blue robot (help-receiver) ran out of battery; its viewpoint was static until the green robot shared its battery with it. We conjecture that the static period of the blue robot might have influenced the participants a lot.

\subsection{Limitation}
First, we considered the task we use in this paper. The background of the videos can expand, besides organizing tables and chairs in a meeting room, there are still a lot of different tasks that could be used. In addition, as sharing the battery with the other one is treated as the altruistic behavior in the current setting, we can not compare the effect between human-appearance agent and robot-appearance agent. 

Second, the limitation of the agent appearance is considered. Many previous works have shown that interacting with human-like agents actually enhances the prosocial behavior of participants \cite{r19}\cite{r2}\cite{r25}\cite{r15}. Because of the particularity of our task setting, we did not compare the difference between robot-appearance virtual agent and human-appearance virtual agent. In the further study, we would like to conduct the task setting which can not only matching to robot but also to human.

Third, the limitation of the scoring of altruism is considered. Engel et al.\cite{r48} have found that most dictator games change depending on the experiment basing on the result of a meta-analysis of the dictator game, so changes in the question setting of the game may cause text dependency. Initially the players of dictator game are humans, but in this experiment we are asking not only for humans but also for robots. This can confuse some participants as to what the money allocated to the robots in dictator game means, or who exactly is the beneficiary. With these considerations in mind, constructing a suitable questionnaire to test the altruism to robots and even can test both altruism to humans and robots are something that should be explored in the future.

Fourth, the limitation of measuring only altruism is considered. As the prosocial behavior is a multidimensional concept\cite{r29}\cite{r30}, the effect of perspective-taking should not only be measured on altruism but also other concepts.

Finally, the results of our study demonstrate the positive effect of perspective-taking on altruistic behavior in a cooperation task. However, in the free-description part after watching the help-provider-view video, a few of participants said that they felt a sense of crisis of losing their job in the future while recognized robots could done many work so well. In addition in the human-human interaction tasks, Pierce et al. have shown that taking the perspective of a competitor might led to more unethical behaviors \cite{r21}. The influence of watching a different robot's viewpoint in a more competitive robot task would be an interesting avenue for future work.

\section{Conclusion}
As one of the method that promoting human prosocial behaviors, perspective-taking showing its effectiveness ranging from traditional perspective-taking task to the mediated perspective-taking tasks. The mediated perspective-taking tasks(e.g., text, video, virtual reality, acted demo) perform better than the traditional ones because of it contain more information(visual, auditory, and even haptic). Previous studies have focused on using perspective-taking for promoting participants' prosocial behaviors through human-like virtual agents. 

In this paper, we investigated how taking the perspective of different robots in a robot-altruistic task affected the social behavior of participants through video stimulus. In the task, the help-receiver robot ran out of the battery and the help-provider robot came toward to it and shared its battery with the help-receiver robot. We ask one group of participants watched the help-receiver-view video and the other group watched the help-provider-view video and use dictator game to measure participants' behavior changes to two targets-- the other participant and a robot.

Our findings showed that participants watching the help-receiver view exhibited more altruistic behaviors toward a robot than those who watched the help-provider view. We also found that watching two different viewpoints from the two robots in the task did not result in any behavior changes with regard to the other participant. In future work, we will apply our study to immersive virtual reality and clarify the effects of different tasks and relationships.

\end{document}